\documentclass{iaus}
\usepackage{graphicx}

\title[IAU Symposium 275.~~Jets at all scales] %% give here short title %%
{Relativistic jets in Narrow-Line Seyfert 1}

\author[L. Foschini et al.]   %% give here short author list %%
{L. Foschini$^1$, E. Angelakis$^2$, G. Bonnoli$^1$, G. Calderone$^3$, M. Colpi$^3$, F. D'Ammando$^4$, D. Donato$^5$, A. Falcone$^6$, L. Fuhrmann$^2$, G. Ghisellini$^1$, G. Ghirlanda$^1$, M. Hauser$^7$, Y.Y. Kovalev$^{2,8}$, L. Maraschi$^1$, E. Nieppola$^9$, J. Richards$^{10}$, A. Stamerra$^{11}$, G. Tagliaferri$^1$, F. Tavecchio$^1$, D.J. Thompson$^5$, O. Tibolla$^{12}$, A. Tramacere$^{13}$, S. Wagner$^7$}

\affiliation{$^1$ INAF -- Osservatorio Astronomico di Brera, 23807 Merate (LC), Italy\\[\affilskip]
$^2$ Max-Planck-Institut f\"ur Radioastronomie, 53121 Bonn, Germany\\[\affilskip]
$^3$ University of Milano Bicocca, 20100 Milano, Italy\\[\affilskip]
$^4$ INAF -- IASF-Palermo, 90146, Palermo, Italy\\[\affilskip]
$^5$ NASA Goddard Space Flight Center, Greenbelt, MD 20771, USA\\[\affilskip]
$^6$ Penn State University, University Park, PA 16802, USA\\[\affilskip]
$^7$ Landessternwarte, Universit\"at Heidelberg, K\"onigstuhl, D 69117 Heidelberg, Germany\\[\affilskip]
$^8$ Astro Space Center of the Lebedev Physical Institute, 117997 Moscow, Russia\\[\affilskip]
$^9$ Mets\"ahovi Radio Observatory, FIN-02540 Kylmala, Finland\\[\affilskip]
$^{10}$ California Institute of Technology, Pasadena, CA 91125, USA\\[\affilskip]
$^{11}$ University of Siena, 53100, Siena, Italy \\[\affilskip]
$^{12}$ University of W\"urzburg, 97074, W\"urzburg, Germany\\[\affilskip]
$^{13}$ INTEGRAL Science Data Centre, CH-1290, Versoix, Switzerland \\[\affilskip]
}

\pubyear{2010}
\volume{275}  %% insert here IAU Symposium No.
\pagerange{???--???}
\jname{Jets at all scales}
\editors{G. E. Romero, R. A. Sunyaev \& T. Belloni, eds.}
\begin{document}

\maketitle

\begin{abstract}
Narrow-Line Seyfert 1 (NLS1) class of active galactic nuclei (AGNs) is generally radio-quiet, but a small percent of them are radio-loud. The recent discovery by \emph{Fermi}/LAT of high-energy $\gamma-$ray emission from 4 NLS1s proved the existence of relativistic jets in these systems. It is therefore important to study this new class of $\gamma-$ray emitting AGNs. Here we report preliminary results about the observations of the July 2010 $\gamma-$ray outburst of PMN~J$0948+0022$, when the source flux exceeded for the first time $10^{-6}$~ph~cm$^{-2}$~s$^{-1}$ ($E>100$~MeV). 
\end{abstract}

The recent discovery of variable $\gamma-$ray emission from 4 NLS1s revealed the presence of a third class of $\gamma-$ray AGNs (Abdo et al. 2009a). This poses intriguing questions to the current knowledge of relativistic jet systems and on how these structures are generated. One of these sources, PMN~J$0948+0022$ ($z=0.5846$) is classified a typical NLS1, but it displays also strong, compact and variable radio emission, with inverted spectrum, suggesting the possibility of the presence of a relativistic jet (Zhou et al. 2003). The confirmation came with the detection of high-energy variable $\gamma$ rays by \emph{Fermi}/LAT (Abdo et al. 2009b, Foschini et al. 2010a). A multiwavelength campaign performed in March-July 2009 displayed coordinated variability at all frequencies, thus confirming that the source detected by \emph{Fermi} is indeed the high-energy counterpart of PMN~J$0948+0022$ (Abdo et al. 2009c). 

PMN~J$0948+0022$, the first NLS1 detected at $\gamma$ rays, was soon followed by three more (Abdo et al. 2009a). Their main differences with respect to blazars and radio galaxies are the optical spectrum and the radio morphology, which is quite compact and without extended structures. These characteristics point to systems with relatively low masses of the central black hole ($10^{6-8} M_{\odot}$) and high accretion rates (up to 90\% of the Eddington value), common for NLS1s, but not for blazars or radio galaxies. In addition, since the $\gamma-$ray NLS1s seem similar to blazars, i.e. small viewing angles, there should be a parent population with the jet viewed at large angles (as blazars vs radio galaxies). The first source of this type has been recently found (PKS~$0558-504$, Gliozzi et al. 2010). Therefore, it seems that NLS1s could be a low mass set of systems ``parallel'' to blazars and radio galaxies. 

\begin{figure}[t]
\begin{center}
 \includegraphics[angle=270,scale=0.4]{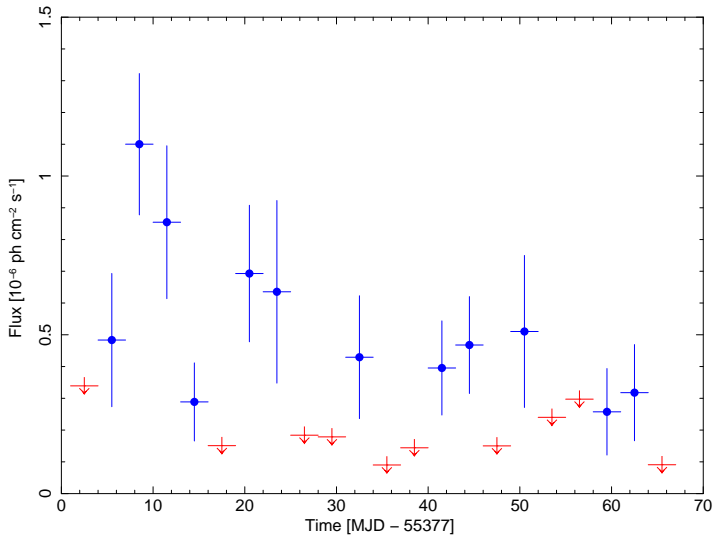} 
 \caption{\emph{Fermi}/LAT lightcurve ($0.1-100$~GeV) of PMN~J$0948+0022$ in July-August 2010, with $3$~days time bin. Blue points are detection with $TS>10$ ($\sim 3\sigma$), while the others are $2\sigma$ upper limits. LAT data have been analyzed as described in Foschini et al. (2010b).}
   \label{fig1}
\end{center}
\end{figure}

A key question was the power released by jets of NLS1s. The early observations and the 2009 MW campaign have shown that the maximum luminosity reached by PMN~J$0948+0022$, the most powerful of these NLS1s, is $\sim 10^{47}$~erg~s$^{-1}$ ($0.1-100$~GeV). On the other hand, blazars can reach greater luminosities ($\sim 10^{49}$~erg~s$^{-1}$ in the case of 3C~454.3, e.g. Foschini et al. 2010b). The question was answered in July 2010, when PMN~J$0948+0022$ underwent a strong outburst (Donato et al. 2010, Foschini 2010c) with a peak flux of $\sim 10^{-6}$~ph~cm$^{-2}$~s$^{-1}$ ($0.1-100$~GeV), corresponding to a luminosity of $\sim 10^{48}$~erg~s$^{-1}$ (Fig.~\ref{fig1}). Even if the source position was too close to the Sun for a full MW campaign, some coverage was obtained. Further details will be available in a forthcoming paper.

\end{document}